\documentclass[runningheads]{llncs}

\usepackage[T1]{fontenc}

\usepackage{todonotes}
\usepackage{graphicx} 
\usepackage{markdown}

\begin{document}

\title{Decoupled Recommender Systems: Exploring Alternative Recommender Ecosystem Designs}

\author{Anas Buhayh\inst{1}\orcidID{0009-0009-7987-7967} \and
Elizabeth McKinnie\inst{1}\orcidID{0009-0002-8721-5700} \and
Robin Burke\inst{1}\orcidID{0000-0001-5766-6434}}
\authorrunning{A. Buhayh et al.}
\titlerunning{Decoupled Recommender Systems (RecSoGood 2024)}
%
\institute{Department of Information Science, University of Colorado, Boulder 
\email{\{anas.buhayh,elizabeth.mckinnie,robin.burke\}@colorado.edu}}

\newcommand{\eliz}[1]{\textcolor{blue}{{\bf [Elizabeth: }{\em #1}{\bf ]}}}

\maketitle

\begin{abstract}
Recommender ecosystems are an emerging subject of research. Such research examines how the characteristics of algorithms, recommendation consumers, and item providers influence system dynamics and long-term outcomes. One architectural possibility that has not yet been widely explored in this line of research is the consequences of a configuration in which recommendation algorithms are decoupled from the platforms they serve. This is sometimes called ``the friendly neighborhood algorithm store'' or ``middleware'' model. We are particularly interested in how such architectures might offer a range of different distributions of utility across consumers, providers, and recommendation platforms. In this paper, we create a model of a recommendation ecosystem that incorporates algorithm choice and examine the outcomes of such a design.
\end{abstract}




\section{Introduction}
An implicit architectural assumption in recommender systems research is the item data/recommendation platform monolith: a single centralized database of all items that can be recommended, a single recommendation algorithm operating over them, and a single centralized database of consumer profiles resulting from user interactions with those items. Fielded systems often contain multiple recommenders operating over different items within the same application: think recommending posts and recommending users to follow in a social media system, or session-oriented and long-term recommendations coexisting in an e-commerce setting. However, these are still monoliths, enabled by a central representation of the user profile and item catalog.

As a personalization technology, recommender systems have historically had a strong focus on the consumers, for whom the experience is personalized, and the evaluation of recommender systems has, appropriately, taken this perspective. In a multistakeholder approach \cite{Abdollahpouri2020}, the field of view is expanded to consider how the recommender system impacts providers of recommended items and others, in addition to item consumers. But even when adopting a multistakeholder lens, the idea of the monolithic recommender remains. We believe this architectural design enforces a ``one-size-fits-all'' recommender system objective on both consumers and providers on the platform. Even in peer-to-peer recommendation, for example \cite{ruffo2009peer}, the user data is moved to the client but we do not see algorithmic diversity as a goal. 

Rajendra-Nicolucci et al. \cite{Rajendra-Nicolucci2023} introduce the \textit{friendly neighborhood algorithm store} model, which emphasizes user choice among third-party algorithms for recommendation designed to meet individual needs. This model creates opportunities for innovation and empowerment for both consumers and providers, fostering healthy competition among algorithm designers. For instance, algorithm designers could develop tools that focus on niche products, ensure the quality of verified content, or create more governable spaces for communities by deploying special content moderation or re-ranking strategies. This work echoes calls from Fukuyama et al. \cite{fukuyama2021save} and others that call for solutions that separate user interface functions from back-end platform operations, especially for social media platforms, an approach they term \textit{middleware}.

This work was also inspired by the work of Yao et al. \cite{yao2023bad}, who used simulations to show that recommender ecosystems tend towards low utility for consumers interested in niche content because such systems incentivize providers to alter their content in the direction of mainstream, popular material. 

In this paper, we use simulations to study the implications of the friendly algorithm store -- or generally, a collection of third-party recommender systems -- to address one of the problems identified in \cite{Rajendra-Nicolucci2023} and confirmed in \cite{yao2023bad}, which is the inability of a single dominant algorithm to meet the needs of all users. To the best of our knowledge, no prior work has simulated recommender ecosystems with decoupled architectures. Our research aims to address the following questions:

\begin{itemize}
\item How do multiple recommender systems interact and evolve within an ecosystem, and what insights can we glean from these interactions?
\item In a decoupled ecosystem, how is utility distributed among providers, consumers, and recommender platforms, and what implications does this distribution have for system dynamics and outcomes?
\item How are these properties influenced by domain- and application-specific characteristics such as the long-tail properties of item popularity, the specific properties of different recommendation algorithms, and the distribution and stability of consumer interests?
\end{itemize}

\begin{figure}
    \centering
    \includegraphics[width=0.9\linewidth]{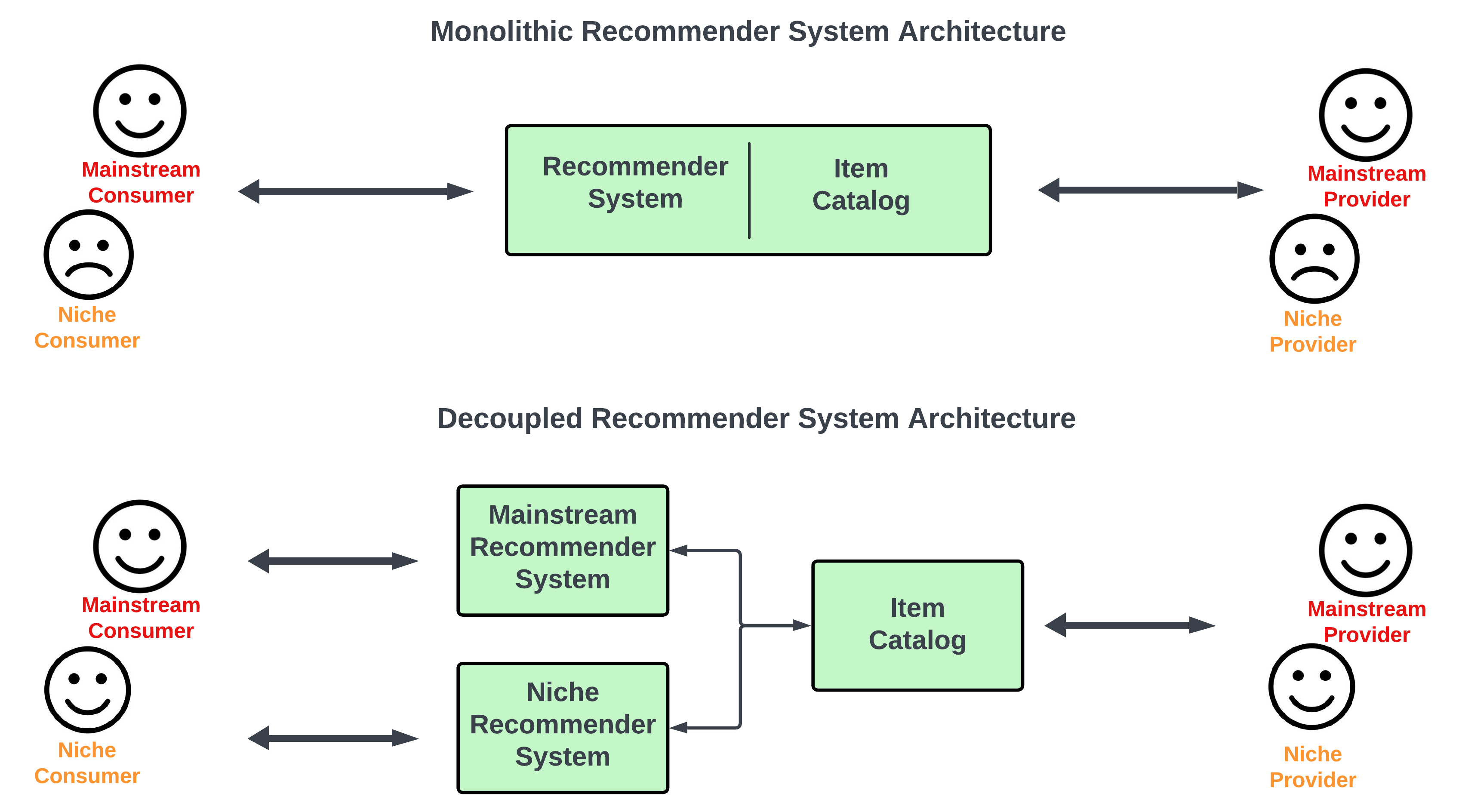}
    \caption{The monolithic vs decoupled recommender system architecture. In the decoupled scenario, consumers can choose among recommender systems and both consumers and providers benefit.} 
    \label{fig:experiment}
\end{figure}

To begin to explore these questions, we developed the Simulator for MOdular Recommendation EcoSystems (SMORES) \footnote{https://github.com/AnasBuhayh/smores} that models the complex interactions between consumers, providers, and recommender platforms in a recommender ecosystem, focusing on how utility is distributed among these three classes of stakeholders. Figure~\ref{fig:experiment} shows a schematic depiction of the different ecosystem structures that we consider: a typical monolithic configuration versus a decoupled one. In this paper, we present our preliminary investigations using SMORES to study the impact of consumers switching among recommender platforms. Our findings demonstrate that introducing a specialized niche recommender system can significantly enhance the utility for both consumers interested in niche items and the providers who produce them. The key contributions of our work are as follows:

\begin{enumerate}
\item We describe SMORES and show how it enables the study of the dynamics of decoupled recommender ecosystems.
\item Using SMORES, we construct a simple recommender choice scenario to mimic the interaction between a consumer and multiple recommender platforms.
\item Under this scenario, we demonstrate that recommendation platform choice provides better outcomes for both niche providers and niche consumers.
\end{enumerate}

\section{Related Work}

Our work contributes to the trajectory of exploring and evaluating multistakeholder recommendation systems, focusing on all groups and individuals interacting with and affected by these systems \cite{burke_multisided_2017,Abdollahpouri2020}. The multistakeholder recommendation concept delineates the ecosystem into three primary components: \(consumers\), who consume the recommendations; \(providers\), who produce the items to be recommended; and the \(system\), encompassing the recommendation algorithm and the platform operating it \cite{Abdollahpouri2020}. Previous research in this domain has predominantly focused on understanding and addressing fairness and disparities among the diverse populations interacting with recommender systems \cite{Abdollahpouri2019,RanjbarKermany2021,Wu2022,Smith2023}. Building on this foundation, our work examines the question of recommender algorithm choice from a multistakeholder perspective, looking at how different classes of consumers and providers are impacted by this architectural configuration.

In their simulated experiment, Yao et al. \cite{yao2023bad} demonstrate that traditional top-k recommendations can effectively support social welfare for content creators (providers) when these creators adapt their strategies and content to align with market demands. However, we contend that this approach may lead to an unhealthy, homogeneous market and impose costs on providers who may find it challenging to chase popular tastes by constantly adapting their output. These incentives serve to marginalize niche providers and consumers, further isolating them from the mainstream \cite{choi2023creator}.

Research on recommender systems fairness has shown that these systems often exhibit disparate treatment of providers, primarily driven by popularity bias inherent in collaborative filtering algorithms \cite{Abdollahpouri2019a,Abdollahpouri2019b}. Such biases can create barriers to entry for new providers \cite{Gope2017}, thereby impacting the inclusivity and diversity of recommender ecosystems. Notably, many businesses and individuals rely on platforms housing recommender systems as their primary source of income \cite{Dalal2023,AlvarezdelaVega2021}. Additionally, recommender systems serve as essential content moderation tools on social media platforms \cite{Gillespie2022}. While these moderation techniques effectively reduce the visibility of harmful content and protect providers from abusive comments, they also risk suppressing marginalized voices and limiting the reach of providers' content \cite{Kingsley2022,Haimson2021}. The diverse roles of recommender systems underscore the complex interplay between algorithmic decision-making, platform governance, and societal equity.

Our work is grounded in the principles of the ``pluriverse'' – a concept delineating a diverse world composed of a myriad of smaller worlds, and "the very small online platforms," as well as "the friendly neighborhood algorithm store," as articulated in \cite{Rajendra-Nicolucci2023}. While these concepts have traditionally been associated with Decentralized Online Social Networks, often referred to as the Fediverse \cite{LaCava2021}, this architectural approach emphasizes decentralized and community-driven governance, albeit with accompanying challenges \cite{Bustamante2023}. In our adaptation, we integrate recommender systems into this decentralized framework, thereby empowering consumers to exert greater control over the recommendation algorithms shaping their online interactions.

This work also has a close connection to the \textit{middleware} proposal from \cite{fukuyama2021save}, which envisions recommender systems as a form of middleware that sit on top of existing social media platforms. They envision this as a new dimension of commercial competition, describing it as \textit{``a competitive layer of new companies with transparent algorithms [who] would step in and take over the editorial gateway functions currently filled by dominant technology platforms whose algorithms are opaque.''}

\section{Methods}

\subsection{Simulation Architecture}

We developed the SMORES simulation architecture, inspired by RecSim \cite{ie2019recsim}, to model different versions of a multistakeholder recommender ecosystem shown in Figure~\ref{fig:experiment}. Key elements of this model are our representations of an item $i \in I$ where $I$ is the set of all items; consumer $j \in J$ where $J$ is the set of all consumers; recommender system $k \in K$ where $K$ is the set of all recommender algorithms; and provider $v \in V$ where $V$ is the set of all providers.

\begin{figure}
    \centering
    \includegraphics[width=0.9\linewidth]{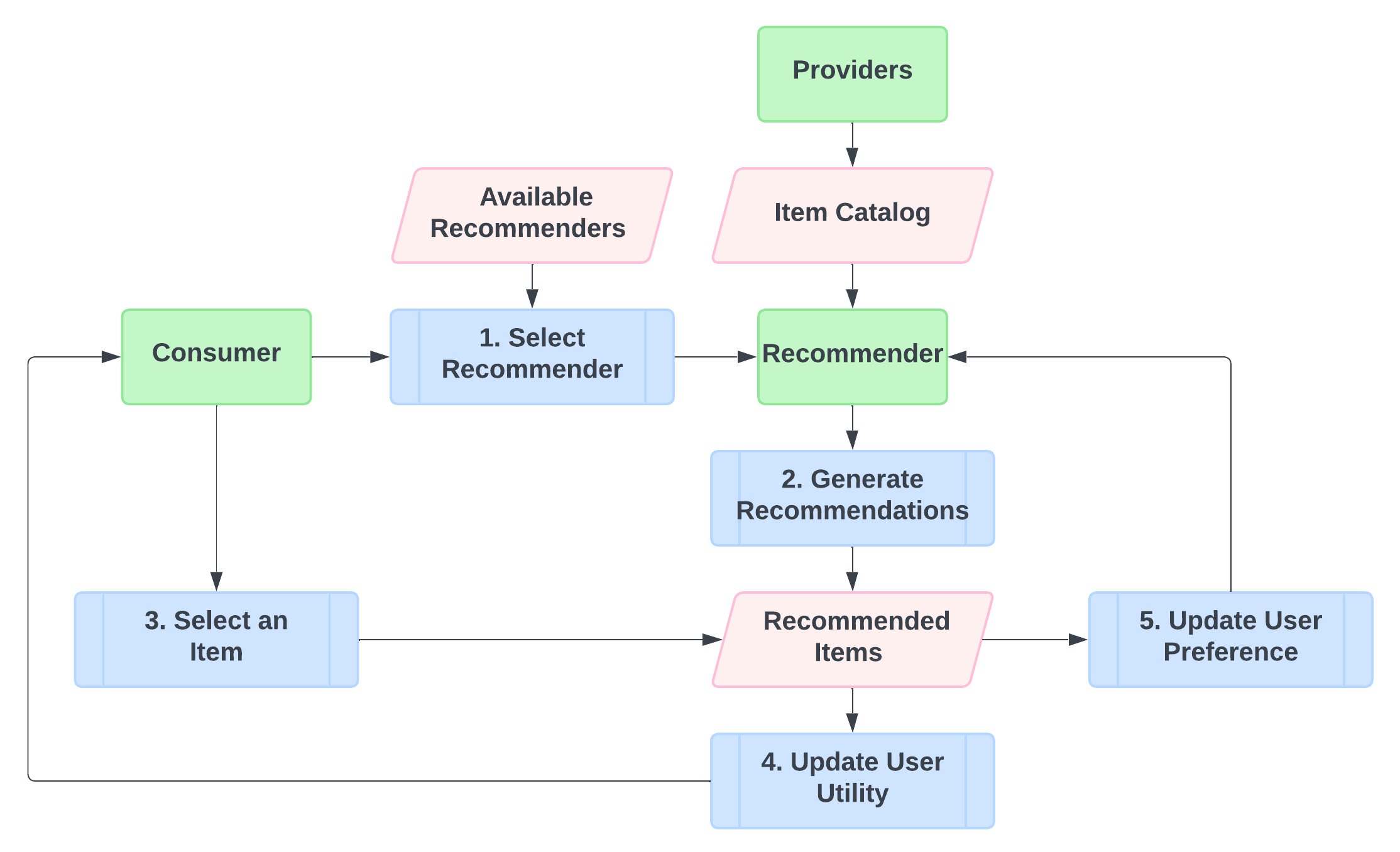}
    \caption{The simulation environment process highlighting the stakeholders in green, processes in blue, and the inputs/outputs in pink.}
    \label{fig:simulation}
\end{figure}
\vspace{-20pt}

As shown in Figure~\ref{fig:simulation}, the basic operation of the simulation involves a consumer $j$ choosing a recommender system $k$ with which to interact, consulting the recommender, being delivered a list of recommended items $[i_0..i_n] = \ell_{j,k}$, and then selecting a particular item $\bar{i}_{j,k}$. In our model, the consumer can only consult one recommender per iteration. We also assume that there is some inertia in consumers' recommender algorithm choices and they update their algorithm choice only periodically, not every time they seek a recommendation. The algorithm produces recommendation lists from the items made available to it by the providers. For these experiments, we assume that all items are available to all recommenders. A sequence in which every user obtains recommendations is termed a \textit{day} and 30 such sequences are a \textit{cycle}. Consumers update their recommender preferences at the end of each cycle. 

Consumers choose an item every time a recommendation list is produced and they obtain utility from that recommendation -- for example, by watching the recommended movie they chose. Providers obtain utility when their items are shown on recommendation lists and when they are selected or clicked on by consumers. The recommender systems also obtain utility in our simulation, but we do not include an analysis of recommender utility in this paper for reasons of space. We detail the simulation calculations below. 

\subsection{Selection Models}

\subsubsection{Recommender Selection Model}

\label{sec:rec-selection}

We experiment with two different recommender selection models. The first is a simple threshold-based model. In this model, the consumer computes their utility towards a recommender system over the entire experiment so far and compares it to a threshold $\tau$. The utility score for the recommender system is computed with a recency bias parameter \( \beta \), which assigns more weight to recent interactions. Let $u_{j,\ell,k}$ be the utility that consumer $j$ associates with a given recommendation list $\ell$ from recommender $k$, and let $U_{j,k}$ be the utility calculated for recommender $k$ so far. An updated $U'_{j,k}$ is computed as:

\begin{equation}
U'_{j,k} = \frac{U_{j,k} \times \beta + u_{j,\ell,k}}{1 + \beta}
\label{eq:user-recommender-utility}
\end{equation}

If $U'_{j,k}$ falls below $\tau$ at the end of a given cycle and there is another recommender to switch to, the consumer will choose the other recommender for the next cycle. 

The second selection model uses the well-known Upper Confidence Bound (UCB) algorithm from the multi-armed bandit literature \cite{slivkins2019introduction}. The UCB formula with decay is expressed as follows: $\text{UCB}_{j,k} = U_{j,k} + \frac{\sqrt{2 \log \left(t \right) / \text{n}_{j,k}}}{1 + t}$, where \( U_{j,k} \) represents the utility score associated with recommender \( k \) based on historical interactions with the recommender and updated using the formula above, \( t \) denotes the current day of interaction, and \( \text{n}_{j,k} \) is the number of times recommender \( k \) has been selected by consumer $j$. This approach ensures that the exploration rate diminishes over time, allowing for more focused exploitation of the best-performing recommenders. At the end of each cycle, consumer $j$ will select the recommender with the greatest $\text{UCB}_{j,k}$ value.

\subsubsection{Item Selection Model}


When a recommendation list $\ell$ is presented to the consumer, the utility $u_{j,i}$ of each item $i \in \ell_{j,k}$ is computed as the dot product of the consumers’ feature interests and the selected item's features: $f_i \cdot p_{j}$ 
where $f_i$ is the feature vector associated with item $i$ and $p_{j}$ is the preference over the same features associated with consumer $j$. This utility score is then normalized by the size of the feature set and converted into selection probabilities using the softmax function:

\begin{equation}
\pi_{j,i} = \frac{\exp(u_{j,i} - \max(u_\ell))}{\sum_{x \in \ell} \exp(u_{j,x} - \max(u_\ell))}
\end{equation}

The selected item $\bar{i}$ is chosen at random using the $\pi_i$ values as probabilities. This item is identified as the \textit{clicked} item for provider utility calculation. Note that a consumer \textit{must} choose an item each day. This design decision implies that even if an entire set of recommendations does not match the user preferences, some selection will be made and some provider will gain utility from that choice. In future work, we plan to explore decision models in which consumers have the null option of not clicking on any item.

\subsection{Utility Models}

\subsubsection{Consumer Utility Model} 

Although the consumer always selects a single item in each recommendation list, their utility is computed from the recommendation list as a whole using the feature-based model discussed above: the utility of a single item is computed as the dot product of the user's feature preferences and the item's features, normalized by the feature count. The average of these values is computed over all of the items in the list, yielding the overall list utility $u_{j,\ell,k}$. The overall utility of a particular recommender is updated as given in Equation~\ref{eq:user-recommender-utility}. 

\subsubsection{Provider Utility Model}

At each cycle, the provider’s utility is updated based on fees charged by the connected recommender systems. This fee structure includes a display fee (or utility cost) $\delta_d$, charged when the provider's item appears on a recommendation list, and a click fee $\delta_c$, charged when a consumer selects the provider's item. There is also utility associated with having one's item appear on a recommendation list, $\phi_d$, and having one's item clicked on, $\phi_c$. The total utility for provider $v$ relative to recommender $k$ is calculated as: $u_{v,k} = (\phi_d - \delta_d) \times n_{d,v,k} + (\phi_c - \delta_c) \times n_{c,v,k}$ where $n_{d,v,k}$ is the number of times that an item from provider $v$ has been displayed by recommender $k$ and $n_{c,v,k}$ is similar but for clicks. 

\subsubsection{Recommender System Utility Model}

The utility accrued by each recommender algorithm is computed based on the  fees assessed on providers. We do not focus on or report recommender system utility in this paper but we expect that, in a decoupled recommendation architecture, different recommendation platforms might have different business models and we plan to explore a diversity of such options in future work.

\subsection{Dataset}

We chose to anchor our simulated results on data from a real-world dataset, the well-known MovieLens 100k dataset \cite{harper2015movielens}. The items are movies and the features for each movie are the movie's associated genre labels. 

We sampled 600 users from the dataset as our consumers and, calculated consumer preference vectors based on the frequency of genres in their preferred items. However, to highlight the concept of \textit{niche consumers} and \textit{niche items}, we manipulated these profiles as follows. We selected the \textit{Western} genre, one of the genres less prevalent in the dataset, as our niche genre. For the 10\% of users with an affinity for this genre, we enhanced their affinity by multiplying the \textit{Western} feature by 4, while reducing the features for other genres proportionately. We did the opposite for our \textit{mainstream} users, shrinking their interest in the \textit{Western} genre by a factor of $1/4$. 

\subsection{Providers}

We generated 10 providers: nine of these providers each offered a random, equal sample of items (n=100) to ensure that our focus remained on the impact of content type rather than quantity. The tenth provider specialized exclusively in the niche items, offering a collection comprised of movies from the \textit{Western} genre.

\subsection{Recommenders}

We implemented a content-based \textit{Mainstream} recommender. The recommender uses a naive Bayes model to update consumer preferences over genres based on clicks and then populates the recommendation list with the most popular movies of these genres. To explore outside of the most popular movies of these genres, the recommender uses an exploration probability (\%20) to randomly select movies of the same genres the consumer historically interacted with to create their recommendation list. We plan to explore additional recommendation algorithms in future work, including collaborative techniques. The \textit{Niche} recommender was implemented similarly, however, concentrating exclusively on the \textit{Western} genre.

\subsection{Experiments}

We conducted three experiments using this simulation framework:

\begin{enumerate}
    \item \textbf{Single recommender}: A single recommender (the \textit{Mainstream} recommender) provides recommendations to all consumers: the standard monolithic design.
    \item \textbf{Decoupled recommenders with threshold}: We introduce the \textit{Niche} recommender in parallel with the \textit{Mainstream} recommender; consumers switch between recommenders using the threshold criterion described above.
    \item \textbf{Decoupled recommenders with UCB}: The same pair of recommenders as above provide recommendations, but the consumers use the upper confidence bound (UCB) criterion to decide between recommenders.
\end{enumerate}

Other simulation parameters include: days per cycle, 30; cycles per experiment, 60; recommendation list size, 5; click utility ($\phi_c$), 0.4; display utility ($\phi_d$), 0.1; click fee ($\delta_c$), 0.1; display fee ($\delta_d$), 0.01;
switching threshold ($\tau$), 0.04.

\section{Results}

\subsection{Single Recommender Experiment}
\begin{figure}
    \centering
    \includegraphics[width=0.8\linewidth]{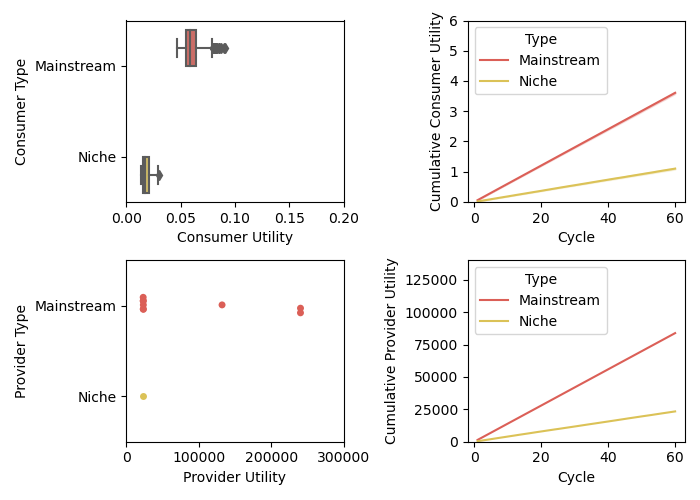}
    \caption{Single recommender experiment: Consumer and provider utility, distribution and cumulative results. The top left box plot shows consumer utility on the last day, while the bottom left strip plot shows provider utility on the last day. The right column line plots depict cumulative consumer utility and provider utility over all cycles.}
    \label{fig:exp1_results}
\end{figure}

The results for the first experiment are shown in Figure~\ref{fig:exp1_results}. The figures at the top provide two different representations of consumer utility, broken down by mainstream and niche consumers. The box plot on the left shows the distribution of utility during the \textit{last day} of the last cycle, representing the final state of the simulation. We see that niche consumers have a much lower utility than the more mainstream consumers of the system. The figure at the right confirms this outcome over the course of the experiment. It shows the average consumer utility (cumulatively) for each type of consumer and again we see that mainstream consumers have much higher utility.

For providers (lower part of the figure), the story is very similar. The single niche provider has much lower utility both at the end of the simulation snapshot and the cumulative plot. 

\subsection{Threshold-based Switching Experiment}
\begin{figure}[tbh]
    \centering
    \includegraphics[width=0.8\linewidth]{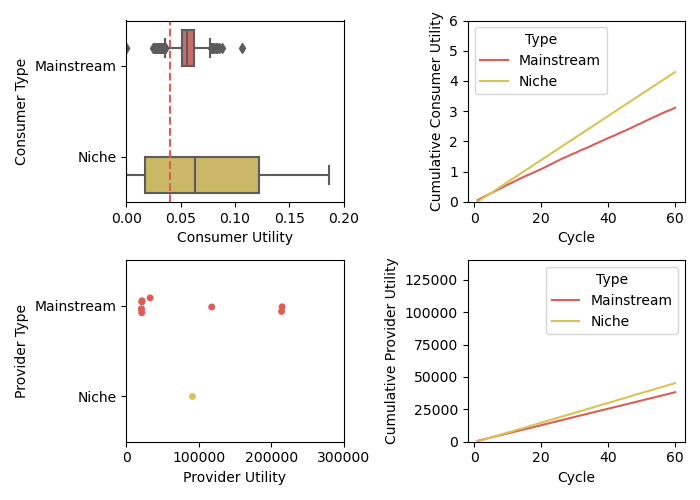}
    \caption{Decoupled recommender with threshold-based switching experiment. The dashed line in the top left plot represents the switching threshold. The provider utility is the sum of the two recommenders' utility for that provider.}
    \label{fig:exp2_results}
\end{figure}

Figure~\ref{fig:exp2_results} is in the same format as Figure~\ref{fig:exp1_results} but for the threshold-based switching experiment. At the start of the experiment, all consumers were initially connected to the \textit{Mainstream} recommender. Consumers will switch between recommenders if their utility falls below the threshold $\tau = 0.04$ at the end of a cycle. 

The decoupled recommender configuration provides an avenue for niche consumers dissatisfied with the \textit{Mainstream} recommender. We see that the utilities are much improved for niche consumers and the mainstream consumers are relatively unaffected. 

The story for providers is a bit different. We see more equity across the providers, but the mainstream providers lose utility because their items are not being forced on niche consumers who are not that interested in them, and now that the niche provider has an avenue to reach interested consumers, they can become one of the more profitable providers.

\subsection{UCB Switching Experiment}
The third experiment was very similar to the second but the switching criterion is based on the upper confidence bound (UCB) technique described in Section~\ref{sec:rec-selection}. This technique allows consumers to select the best model for them, based on their own experiences, rather than having to wait until their experience is so bad that it hits the threshold before switching. 

\begin{figure}[tbh]
    \centering
    \includegraphics[width=0.8\linewidth]{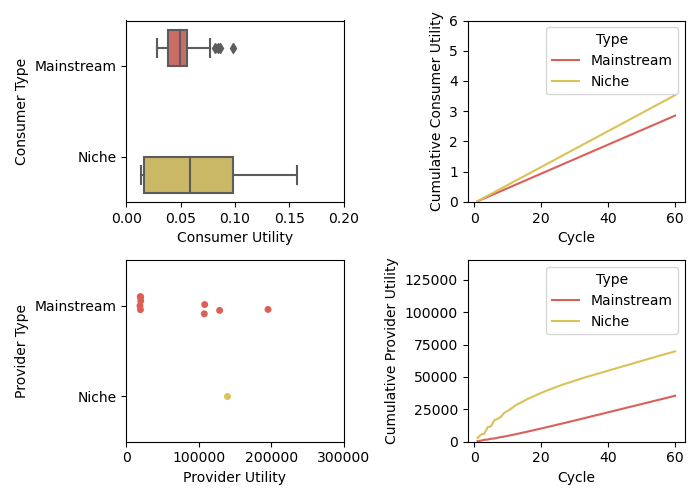}
    \caption{Decoupled recommender with UCB-based switching. }
    \label{fig:exp3_results}
\end{figure}

The results from this experiment are shown in Figure~\ref{fig:exp3_results} and are similar to the threshold-based switching experiment results for user utility. A bigger difference is shown in the utility for the \textit{Niche} provider. This is most likely because the exploration mechanism of UCB means that more \textit{Mainstream} consumers end up in the \textit{Niche} recommender for at least one cycle, and while there, are required to click on (and can \textit{only} click on) the niche provider's items. We note that the cumulative utility curve becomes linear (and more like that of Experiment 2) after about 20 cycles. Presumably by this point, the exploration phase is more or less complete, \textit{Mainstream} consumers stay where they are, and the \textit{Niche} provider is receiving utility just based on Niche consumers' clicks. 

\vspace{-20pt}

\subsection{Comparing experiment results}
\begin{figure}[tbh]
    \centering
    \includegraphics[width=1\linewidth]{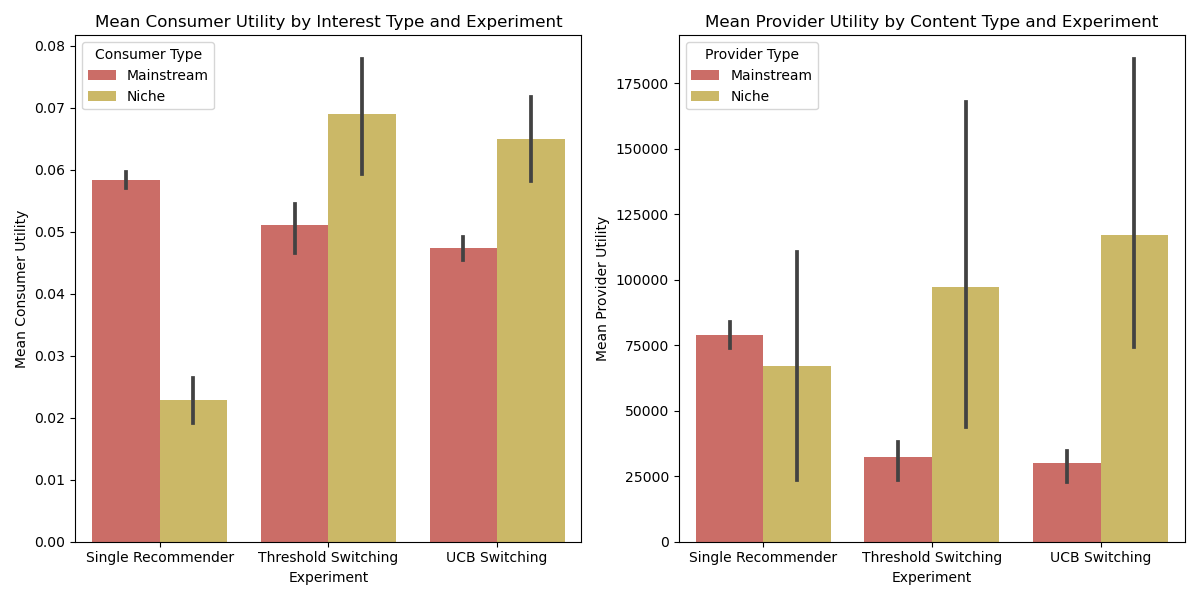}
    \caption{Mean utility scores on the last day for five random seeds of the three experiments for providers and consumers.}
    \label{fig:final_img}
\end{figure}

To further validate our results and provide a contrast between the three experimental conditions, we ran our three experiments on five random seeds and recorded the mean per-day utilities for the niche and the mainstream consumers and for the niche and mainstream providers. See Figure~\ref{fig:final_img}. As the individual experiment results have indicated, niche consumer utility is greatly enhanced by the decoupled design, and the type of switching criteria (threshold-based or UCB-based) makes little difference. The mainstream consumers do see a bit of utility loss under these conditions, probably because they switch to the niche recommender and do not get good results. 

The provider story is different. Niche providers gain in the decoupled environment but within the variability of the simulation, it is hard to say how significant this increase is in general. The mainstream providers do lose out because they have fewer captive consumers than in the monolithic case. We expect this difference to be reduced if consumers could opt out of clicking when presented with low-utility items.

\section{Conclusion}

In this paper, we use our recommender ecosystem simulation SMORES to study the concept of \textit{decoupled recommendation}, where recommender algorithms compete for recommendation consumers. We confirm findings from prior ecosystem work, which show that niche consumers and niche providers tend to do poorly in monolithic systems. The decoupled alternative is better for both providers and consumers but is particularly beneficial for niche consumers with marginal costs for mainstream consumers. This work therefore lends support to calls made in policy and governance circles that envision an algorithm marketplace, allowing consumers to choose algorithms to be applied to the platforms that they use.

The work described here represents an initial exploration of this concept with a great deal of additional study that can be performed. We have constructed an artificial setting where niche consumers are quite selective of and mainstream consumers are quite averse to niche items. Also, there is only one niche genre and only one recommender alternative. We expect to relax these assumptions in future work. We have only explored a content-based mainstream recommender and a niche-specific recommender, but there is room to explore additional designs including collaborative recommendation. We are also interested in exploring how robust our findings are relative to changes in the selection models and utility models. 

There are many directions in which to explore different algorithmic utility models. In our current model, the recommenders are free to consumers and provider-supported but all providers are associated with all recommenders. We do not currently model the interaction between the item catalog platform and the recommenders or the variety of business models that recommenders might pursue. This is an important avenue to consider if such decoupled recommender system architectures are to be implemented and supported.

\bibliographystyle{splncs04}
\bibliography{av-references}

\begin{thebibliography}{10}
\providecommand{\url}[1]{\texttt{#1}}
\providecommand{\urlprefix}{URL }
\providecommand{\doi}[1]{https://doi.org/#1}

\bibitem{Abdollahpouri2019b}
Abdollahpouri, H.: Popularity bias in ranking and recommendation. In: Proceedings of the 2019 AAAI/ACM Conference on AI, Ethics, and Society. pp. 529--530. AIES '19, Association for Computing Machinery (2019). \doi{10.1145/3306618.3314309}, \url{https://doi.org/10.1145/3306618.3314309}

\bibitem{Abdollahpouri2020}
Abdollahpouri, H., Adomavicius, G., Burke, R., Guy, I., Jannach, D., Kamishima, T., Krasnodebski, J., Pizzato, L.: Multistakeholder recommendation: Survey and research directions. User Modeling and User-Adapted Interaction  \textbf{30}(1),  127--158 (March 1 2020). \doi{10.1007/s11257-019-09256-1}, \url{https://doi.org/10.1007/s11257-019-09256-1}

\bibitem{Abdollahpouri2019}
Abdollahpouri, H., Burke, R.: Multi-stakeholder recommendation and its connection to multi-sided fairness (July 30 2019). \doi{10.48550/arXiv.1907.13158}, \url{https://doi.org/10.48550/arXiv.1907.13158}

\bibitem{Abdollahpouri2019a}
Abdollahpouri, H., Mansoury, M., Burke, R., Mobasher, B.: The unfairness of popularity bias in recommendation (September 19 2019), \url{http://arxiv.org/abs/1907.13286}

\bibitem{AlvarezdelaVega2021}
Alvarez De La~Vega, J., Cecchinato, M., Rooksby, J.: 'why lose control?' a study of freelancers' experiences with gig economy platforms. In: Kitamura, Y., Quigley, A., Isbister, K., Igarashi, T., Bj{\o}rn, P., Drucker, S. (eds.) Proceedings of the 2021 CHI Conference on Human Factors in Computing Systems. p.~455. ACM, New York (2021). \doi{10.1145/3411764.3445305}, \url{https://doi.org/10.1145/3411764.3445305}

\bibitem{burke_multisided_2017}
Burke, R.: Multisided fairness for recommendation. 2017 Workshop on {Fairness}, {Accountability} and {Transparency} in {Machine} {Learning} ({FAT/ML 2017}) (2017)

\bibitem{Bustamante2023}
Bustamante, P., Gomez, M., Krishnamurthy, P., Madison, M.J., Murtazashvili, I., Palanisamy, B., Palida, A., Weiss, M.B.H.: On the governance of federated platforms. SSRN Scholarly Paper  (August 2023). \doi{10.2139/ssrn.4528712}, \url{https://doi.org/10.2139/ssrn.4528712}

\bibitem{Rajendra-Nicolucci2023}
{Chand Rajendra-Nicolucci, Michael Sugarman, and Ethan Zuckerman}: The three-legged stool: A manifesto for a smaller, denser internet (March 29 2023), \url{https://publicinfrastructure.org/2023/03/29/the-three-legged-stool/}

\bibitem{choi2023creator}
Choi, Y., Kang, E.J., Lee, M.K., Kim, J.: Creator-friendly algorithms: Behaviors, challenges, and design opportunities in algorithmic platforms. In: Proceedings of the 2023 CHI Conference on Human Factors in Computing Systems. pp. 1--22 (2023)

\bibitem{Dalal2023}
Dalal, S., Chiem, N., Karbassi, N., Liu, Y., Monroy-Hern{\'a}ndez, A.: Understanding human intervention in the platform economy: A case study of an indie food delivery service. In: Proceedings of the 2023 CHI Conference on Human Factors in Computing Systems. pp. 1--16. ACM, Hamburg, Germany (2023). \doi{10.1145/3544548.3581517}, \url{https://doi.org/10.1145/3544548.3581517}

\bibitem{fukuyama2021save}
Fukuyama, F., Richman, B., Goel, A.: How to save democracy from technology: ending big tech's information monopoly. Foreign Aff.  \textbf{100}, ~98 (2021)

\bibitem{Gillespie2022}
Gillespie, T.: Do not recommend? reduction as a form of content moderation. Social Media + Society  \textbf{8}(3),  20563051221117552 (July 1 2022). \doi{10.1177/20563051221117552}, \url{https://doi.org/10.1177/20563051221117552}

\bibitem{Gope2017}
Gope, J., Jain, S.K.: A survey on solving cold start problem in recommender systems. In: 2017 International Conference on Computing, Communication and Automation (ICCCA). pp. 133--138 (2017). \doi{10.1109/CCAA.2017.8229786}, \url{https://doi.org/10.1109/CCAA.2017.8229786}

\bibitem{Haimson2021}
Haimson, O.L., Delmonaco, D., Nie, P., Wegner, A.: Disproportionate removals and differing content moderation experiences for conservative, transgender, and black social media users: Marginalization and moderation gray areas. Proceedings of the ACM on Human-Computer Interaction  \textbf{5}(CSCW2),  1--35 (October 2021). \doi{10.1145/3479610}, \url{https://doi.org/10.1145/3479610}

\bibitem{harper2015movielens}
Harper, F.M., Konstan, J.A.: The {M}ovielens datasets: History and context. ACM Transactions on Interactive Intelligent Systems  \textbf{5}(4),  1--19 (2015)

\bibitem{ie2019recsim}
Ie, E., Hsu, C.w., Mladenov, M., Jain, V., Narvekar, S., Wang, J., Wu, R., Boutilier, C.: Recsim: A configurable simulation platform for recommender systems. arXiv  (Sep 2019), \url{http://arxiv.org/abs/1909.04847}

\bibitem{Kingsley2022}
Kingsley, S., Sinha, P., Wang, C., Eslami, M., Hong, J.I.: 'give everybody [..] a little bit more equity': Content creator perspectives and responses to the algorithmic demonetization of content associated with disadvantaged groups. Proceedings of the ACM on Human-Computer Interaction  \textbf{6}(CSCW2),  1--37 (November 2022). \doi{10.1145/3555149}, \url{https://doi.org/10.1145/3555149}

\bibitem{LaCava2021}
La~Cava, L., Greco, S., Tagarelli, A.: Understanding the growth of the fediverse through the lens of mastodon. Applied Network Science  \textbf{6}(1), ~64 (September 1 2021). \doi{10.1007/s41109-021-00392-5}, \url{https://doi.org/10.1007/s41109-021-00392-5}

\bibitem{RanjbarKermany2021}
Ranjbar~Kermany, N., Zhao, W., Yang, J., Wu, J., Pizzato, L.: A fairness-aware multi-stakeholder recommender system. World Wide Web  \textbf{24}(6),  1995--2018 (November 1 2021). \doi{10.1007/s11280-021-00946-8}, \url{https://doi.org/10.1007/s11280-021-00946-8}

\bibitem{ruffo2009peer}
Ruffo, G., Schifanella, R.: A peer-to-peer recommender system based on spontaneous affinities. ACM Transactions on Internet Technology (TOIT)  \textbf{9}(1),  1--34 (2009)

\bibitem{slivkins2019introduction}
Slivkins, A., et~al.: Introduction to multi-armed bandits. Foundations and Trends{\textregistered} in Machine Learning  \textbf{12}(1-2),  1--286 (2019)

\bibitem{Smith2023}
Smith, J.J., Buhayh, A., Kathait, A., Ragothaman, P., Mattei, N., Burke, R., Voida, A.: The many faces of fairness: Exploring the institutional logics of multistakeholder microlending recommendation. In: Proceedings of the 2023 ACM Conference on Fairness, Accountability, and Transparency. pp. 1652--1663. FAccT '23, Association for Computing Machinery (2023). \doi{10.1145/3593013.3594106}, \url{https://doi.org/10.1145/3593013.3594106}

\bibitem{Wu2022}
Wu, H., Ma, C., Mitra, B., Diaz, F., Liu, X.: A multi-objective optimization framework for multi-stakeholder fairness-aware recommendation. ACM Transactions on Information Systems  \textbf{41}(2),  47:1--47:29 (December 21 2022). \doi{10.1145/3564285}, \url{https://doi.org/10.1145/3564285}

\bibitem{yao2023bad}
Yao, F., Li, C., Nekipelov, D., Wang, H., Xu, H.: How bad is top-$ k $ recommendation under competing content creators? In: International Conference on Machine Learning. pp. 39674--39701. PMLR (2023)

\end{thebibliography}
\end{document}